\documentclass[preprint,journal]{vgtc}       

\ifpdf
  \pdfoutput=1\relax                   
  \usepackage{graphicx}                
  \usepackage{soul}                      

  \DeclareGraphicsExtensions{.pdf,.png,.jpg,.jpeg} 
\else
  \usepackage{graphicx}                
  \DeclareGraphicsExtensions{.eps}     
\fi%

\usepackage{microtype}                 
\PassOptionsToPackage{warn}{textcomp}  
\usepackage{textcomp}                  
\usepackage{mathptmx}                  
\usepackage{times}                     
\usepackage{cite}                      
\usepackage{tabu}                      
\usepackage{booktabs}                  
\usepackage{todonotes}              

\usepackage{tabularx}
\usepackage{glossaries}
\usepackage{multirow}

\usepackage{subcaption}
\usepackage{hyperref}

\usepackage{wrapfig}                    
\usepackage{lscape}
\usepackage{rotating}                   

\usepackage{CJKutf8}
\usepackage[utf8]{inputenc} 
\usepackage[T1]{fontenc}
\usepackage{csquotes}                   
\usepackage{arabtex}
\usepackage{utf8}

\newcommand{\ar}[1]{{\color[HTML]{f8766d}{\<#1>}}} 
\newcommand{\arcolor}[1]{{\color[HTML]{f8766d}{#1}}} 
\newcommand{\en}[1]{{\color[HTML]{a3a500}{#1}}}
\newcommand{\fr}[1]{{\color[HTML]{00bf7d}{#1}}} 
\newcommand{\de}[1]{{\color[HTML]{00b0f6}{#1}}}
\newcommand{\zh}[1]{{\color[HTML]{e76bf3}{#1}}} 
\newcommand{\revision}[1]{#1}

\usepackage{xspace}
\newcommand{\etal}{\emph{et al.}\@\xspace}
\newcommand{\ie}{\emph{i.e.}\xspace}
\newcommand{\eg}{\emph{e.g.}\xspace}

\newcommand{\etals}{\mbox{\emph{et~al.}'s }}

\usepackage{enumitem}

\onlineid{1653}

\vgtccategory{Research}
\vgtcpapertype{Evaluation}

\title{\textit{Probablement, Wahrscheinlich, Likely}?
A Cross-Language Study of How People Verbalize Probabilities in Icon Array Visualizations}

\author{No\"{e}lle Rakotondravony, Yiren Ding 
, and Lane Harrison}
\authorfooter{
\item N. Rakotondravony, Y. Ding, and L. Harrison are with Worcester Polytechnic Institute. E-mail: \{ntrakotondravony\,$|$\,yding5\,$|$\,ltharrison\}@wpi.edu\,.
}

\shortauthortitle{Rakotondravony \MakeLowercase{\textit{et al.}}: Likely, Wahrscheinlich, Probablement?
A Cross-Language Study of How People Verbalize Probabilities in Icon Array Visualizationss}

\abstract{
Visualizations today are used across a wide range of languages and cultures. Yet the extent to which language impacts how we reason about data and visualizations remains unclear.
In this paper, we explore the intersection of visualization and language through a cross-language study on estimative probability tasks with icon-array visualizations. 
Across \arcolor{Arabic}, \en{English}, \fr{French}, \de{German}, and \zh{Mandarin}, $n=50$ participants per language both chose probability expressions --- \eg \emph{likely}, \emph{probable} --- to describe icon-array visualizations (Vis-to-Expression), and drew icon-array visualizations to match a given expression (Expression-to-Vis).
Results suggest that there is no clear one-to-one mapping of probability expressions and associated visual ranges between languages. 
Several translated expressions fell significantly above or below the range of the corresponding English expressions.
Compared to other languages, French and German respondents appear to exhibit high levels of consistency between the visualizations they drew and the words they chose.
Participants across languages used similar words when describing scenarios above 80\% chance, with more variance in expressions targeting mid-range and lower values.
We discuss how these results suggest potential differences in the expressiveness of language as it relates to visualization interpretation and design goals, as well as practical implications for translation efforts and future studies at the intersection of languages, culture, and visualization.
Experiment data, source code, and analysis scripts are available at the following repository: \url{https://osf.io/g5d4r/}.
} 

\keywords{Visualization, Cross-Language Study, Icon-Arrays}

\teaser{
  \centering
  \includegraphics[width=0.95\linewidth]{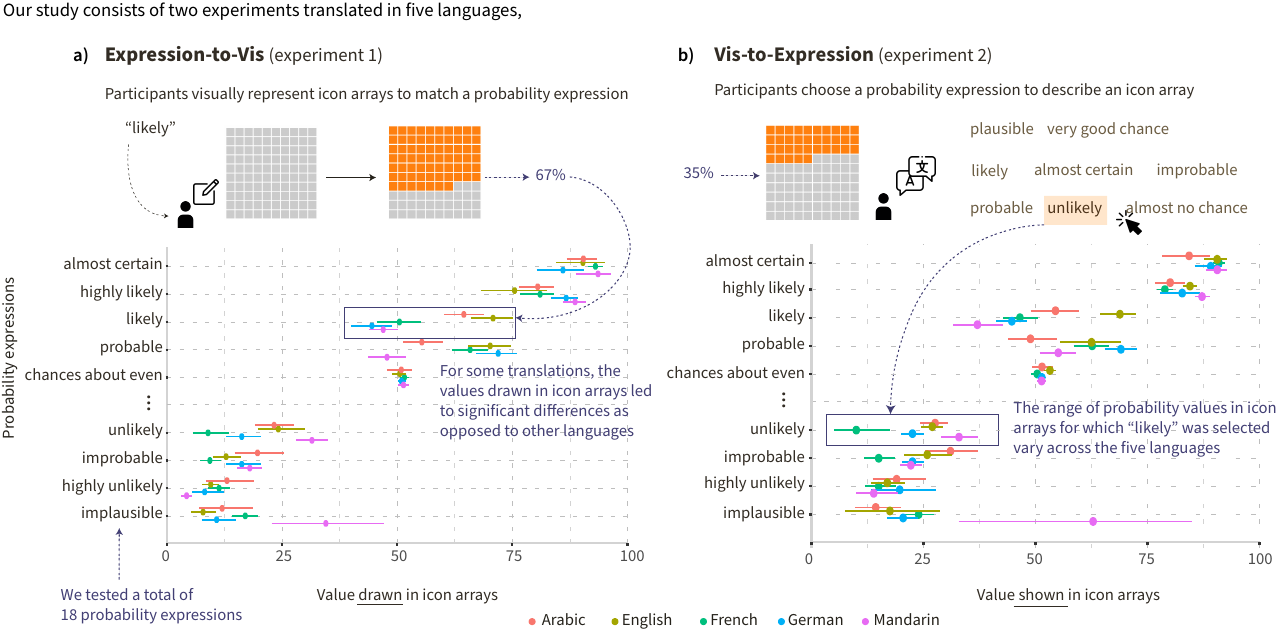}
  \caption{An overview of our cross-language study consisting of two experiments. In (a) \textbf{Expression-to-Vis}, we explore how people draw icon-array visualizations when given particular probability expressions, showing 95\% CIs mean responses across Arabic, English, French, German, and Mandarin. In (b) \textbf{Vis-to-Expression} we investigate how people choose probability expressions to describe icon-array visualizations depicting different values. 95\% CIs show the icon-array ranges to which each expression was assigned by participants, across the studied languages.}
  \label{fig:teaser}
}

\begin{document}
\begin{CJK}{UTF8}{gbsn}
\setcode{utf8}


\maketitle


\section{Introduction}

English remains the dominant language in the study and practice of visualization, but the landscape is changing. 
Creators from diverse languages and cultures are producing more visualizations, in part due to better access to visualization authoring tools and publishing ecosystems.
International newsrooms are producing visualization-laden journalism, and citizens on social media share and discuss visualizations in their native languages.
Global challenges highlight the importance of better understanding of the interplay of language, culture, and visualization. Climate change, pandemics, and misinformation--- all will require a global collective engagement with data to navigate.


Efforts in Human-Computer Interaction show how effects of language and culture might emerge in visualization.
Several HCI studies and design guidelines that focus on WEIRD populations (Western, Educated, Industrialized, Rich, and Democratic) have failed to generalize when considering other languages and cultures \cite{sturm2015weird}.
Other studies have shown that HCI design processes can successfully integrate language and culture as an influence for interaction mechanisms and interface design \cite{evers1998crosscultural}.
More broadly, some research agendas have included multicultural populations and their needs from the outset \cite{reinecke2015labinthewild}.
These findings raise questions for the visualization community: to what extent do studies centering on WEIRD populations generalize to broader global populations?

There are comparatively few studies examining language and culture in visualization.
A notable exception is Kim \etals study of color names across languages \cite{kim2019color}, which found that some languages have more distinct names within certain color ranges, potentially influencing visualization color palette design.
Related to visualization, studies have found that different languages and culture impact the use of color \cite{gibson2017color}, and visual forms of how people represent time \cite{fuhrman2010cross}.

\begin{figure}[tb]
    \centering
    \includegraphics[width=0.6\columnwidth]{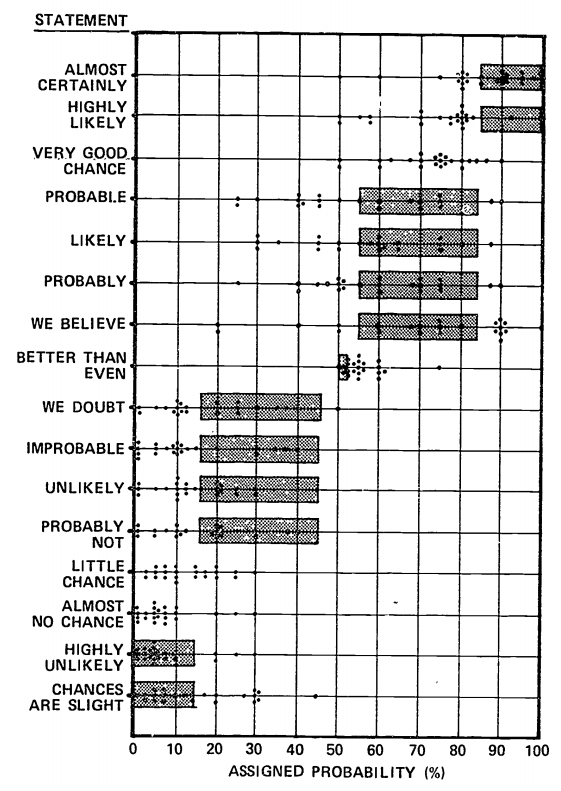}%
    \caption{Results from Kent's survey of 23 intelligence officers. Each dot is a probability assigned to an expression. The shaded areas indicate scale ranges that Kent proposed for the verbal expressions\cite{barclay1977handbook}. For comparison, we superimpose the shaded areas on our results, see \autoref{fig:w2v-95ci-kent}.}
    \label{fig:kent-chart}
\end{figure}

Studies focusing on language and statistics offer a promising means for exploring visualization across languages.
In a widely replicated study, Kent showed that intelligence analysts gave different numerical estimations to the probability expressions used in intelligence reports, such as \emph{``likely"} or \emph{``almost certain"} \cite{kent1964words} (see \autoref{fig:kent-chart}).  
Later studies examined probability expressions across languages.
Renooij and Witteman elicited numerical values for probability expressions with Dutch speakers \cite{renooij1999talking}. 
Doupnik \etal studied how German and English-speaking accountants interpret verbal probability expressions in International Accounting Standards, finding significant differences depending on language \cite{doupnik2003interpretation}. 
Recently, visualization researchers Henkin and Turkay extended similar methodologies to study expressions related to correlation estimation, concluding with an explicit call to examine possible effects across languages \cite{henkin2019words}.


In this paper, we explore the intersection of languages, probability expressions, and visualization.
Extending the expression-to-probability methodologies of Kent \cite{kent1964words}, Renooij and Witteman \cite{renooij1999talking}, and others, we have participants specify values given an expression by drawing icon-array visualizations.
We then invert this procedure by having participants choose expressions for a given icon-array visualization, for a two-part randomized within-subjects study.
We collect expressions from prior studies, resolving issues like phrase asymmetry, ending with a list of $n=18$ base expressions in English. 
To extend to other languages, specifically \fr{French}, \de{German}, \arcolor{Arabic}, and \zh{Mandarin}, we recruit native speakers in each language for a collaborative translation activity, using inter-coder agreement measures to finalize a set of translations.
Using the crowdsourcing platform Prolific, $n=250$ participants ($n=50$ native speakers for each language) completed both Vis-to-Expression and Expression-to-Vis sections.


Results suggest that people vary in how they visualize a given probability expression, with differences both within and across languages. 
Across languages, participants appear to agree more (\ie the response ranges are tighter) when given expressions that indicate higher and lower probability values, such as \emph{very good chance} and \emph{highly unlikely}.
Exceptions exist between languages, however, with some expressions producing substantially different value ranges from corresponding expressions in other languages, see \autoref{fig:w2v-95ci-kent}.
People also vary in how they choose expressions when given an icon-array visualization.
In \arcolor{Arabic} for example, participants chose 15/18 possible expressions when given an icon-array depicting a 40\% chance, compared to 7/18 expressions for \zh{Mandarin}-speaking participants, see \autoref{fig:v2w-scatterplot}.
Additional analyses between experiments reveal differences in elicitation method, where people across languages tended to draw values for a given expression that were more extreme, while less extreme values were common when expressions were chosen for a given icon-array, see \autoref{fig:compare-exp}.

\revision{Taken together, the experiments and results reveal substantial differences in the expressiveness of translated language as it relates to how people interpret visualizations.}
We discuss these findings, and how such differences may impact aspects of the visualization design process, particularly as it relates to communication or visualization translation efforts.
Among other findings, we contribute: 
\vspace{-0.35em}
\begin{itemize}[noitemsep]
    \item Evidence of no clear mapping between drawn visualizations and probability expressions across languages, suggesting cross-language differences, see Figure \ref{fig:w2v-95ci-kent}.
    \item Results suggesting that different languages can exhibit varying degrees of expressiveness for associated icon-array visualizations, see Figure \ref{fig:v2w-scatterplot}.
    \item Cross-language experiment materials in 5 languages and datasets reflecting judgments of $n=250$ participants, including 4,500 \emph{Expression-to-Vis}, and 4,750 \emph{Vis-to-Expression} judgments.
\end{itemize}

\section{Background}

In crafting an experiment targeting estimative probability expressions spanning multiple languages, we draw on methodologies from studies on statistics and language, as well as considerations from cross-cultural studies in the HCI community.
For design choices related to the icon-array visualizations, we refer to several visualization studies using icon-arrays in various contexts.

\begin{figure*}[ht]
    \centering
    \includegraphics[width=0.87\textwidth]{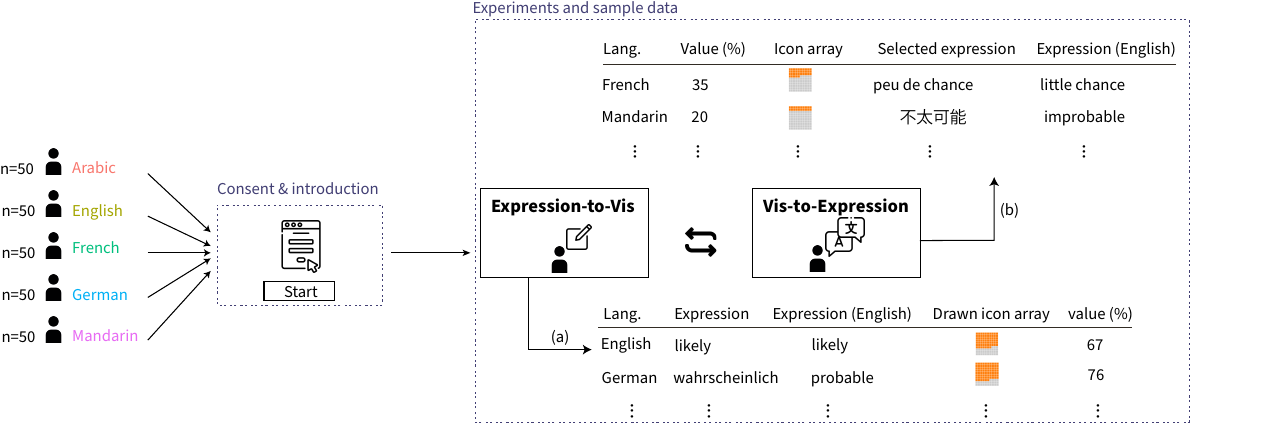}
    \caption{The methodology that we follow in our studies. After signing the consent form and viewing an introduction, participants start the study with either Expression-to-Vis or Vis-to-Expression. \textbf{(a)} and \textbf{(b)} point respectively to sample data sets collected from the two experiments. 
    }
    \label{fig:methodology}
\end{figure*}

\subsection{Probability Expression Interpretation}
Numerical formats can facilitate probability comparison \cite{wallsten1986measuring}. 
Yet because probability is not understood in the same way by everyone, studies have shown that numbers can provide illusory precision \cite{budescu1985consistency}. 
Other studies have shown that people sometimes prefer to communicate uncertainty using verbal expressions in conversation \cite{wintle2019verbal}.
These verbal expressions of probability, however (\eg ``\emph{highly likely, probable}"), can be interpreted differently. 
In a survey of 23 intelligence officers, Kent found variation in the numerical values and ranges that participants assigned to probability expressions that were commonly used in intelligence reports \cite{kent1964words}. 
Kent's work is an early study of the interpretation of probability expressions, and highlighted the uneven relationship between the meaning that a communicator intends and the meaning that the audience may perceive (see \autoref{fig:kent-chart}).
We aim to see if this effect can be found for visualizations of similar values, and whether different patterns are found in languages beyond English.

Empirical studies on the numerical estimation of probability expressions have used elicitation methods comprising word-to-number translations \cite{budescu1985consistency}, number-to-word conversion \cite{reagan1989quantitative}, and rank ordering of expressions \cite{renooij1999talking, mosteller1990quantifying}. 
In these studies, probability expressions are generally studied by giving people probabilistic outcomes for specific scenarios. 
Results from these studies have isolated several potential factors that may impact how people understand, assess, and communicate probabilistic data. 
For example, the combination of verbal and numerical formats like percentage, frequency or numerical range, have been shown to aid peoples' probabilistic reasoning \cite{wintle2019verbal, budescu2014interpretation}.
Other studies have explored factors related to culture and language. 
Doupnik and Richter find that German accountants' interpreted probability expressions in international accounting standards as reflecting significantly lower values than that of their American counterparts \cite{doupnik2004impact}. 
Follow up studies have speculated that this may reflect differences in cultural values where German accountants express more conservatism and stronger risk-avoidance \cite{bocklisch2013doyoumean, doupnik2006influence}.

\subsection{Uncertainty Visualization and Icon-Arrays}
Visual depictions of probability are widely considered to be effective in communicating uncertainty, aiding audiences of different backgrounds to improve decisions, trust and judgment \cite{padilla2020uncertainty, spiegelhalter2011visualizing}. 
Today, uncertainty visualization is widely studied and applied in both scientific domains \cite{brustrenck2013communicating} and in communication with general audiences \cite{kay2016whenish}. 
One of the most common approaches in uncertainty visualizations implements \emph{frequency framing}, in which the probabilistic information are displayed in frequency or ratio format \cite{schapira2001frequency}. 
In a frequency-based representation, the chance of occurrence of an event is shown as a part-to-whole proportion, considered to align better with how people naturally think of probability \cite{gigerenzer1995improve, till2014fostering}. 
Studies have found that uncertainty visualizations using frequency formats tend to be effective in communicating risks, especially for people with low numeracy \cite{zikmundfisher2014blocks, padilla2020uncertainty, galesic2009using, garciaretamero2009communicating}. 

The icon-array is a common visualization type that implements frequency framing.
Icon-arrays typically include one shape (or icon)  repeated a number of times, with some of the shapes colored or otherwise marked to represent a proportion (\eg 35/100).
Several studies have shown that icon-arrays are an effective method for communicating risk, such as simple ratio-based probability values \cite{padilla2021review}. 
This part-to-whole representation of proportion reflects the frequency of events and chances, providing a visual affordances for audiences with different statistical and visualization experience to grasp.
Studies have identified potential additional benefits of icon arrays including 
increased accuracy risk estimation tasks \cite{okan2015improving}, reduced denominator neglect \cite{garciaretamero2009communicating}, and better understanding of medical risk severity \cite{galesic2009using}. 


In this study, we identify a direct connection between probability expressions from studies that focus on statistics and language, and icon-array visualizations.
Building on icon-array designs used in prior work, we use $10x10$ icon arrays and similar encoding methods, following studies by Kreuzmair \etal \cite{kreuzmair2016high}, Bancilhon \etal \cite{bancilhon2019gamble}, Garcia-Retamero \etal \cite{garciaretamero2009communicating}, and Galesic \etal\cite{galesic2009using}.

\subsection{Visualizations, Text, and Language}
While there is little prior work in visualization across languages and culture, topics of visualizations and text or visualizations and natural language provide perspectives that inform the current work.
Verbal, text-based answers visualization tasks are a common methodology, from Cleveland and McGill's graphical perception experiment where people specify a text-based answer \cite{cleveland1984graphical}, to open-ended conversational methodologies such as Peck \etal's study of visualization perceptions in a local farmer's market \cite{peck2019data}.
Visualization is also considered in context with the text that surrounds it, for example Ottley \etal combined text and visualizations for Bayesian reasoning tasks \cite{ottley2016improving}.
Kong \etal show that the text surrounding a visualization impacts how people engage with the visualization itself \cite{kong2018frames}.

The alignment between the perception of data through visualization and the language used to talk about the data has been sought by researchers who investigate the expressiveness of data visualizations. In their study of the ``Words of Estimative Correlation’', Henkin and Turkay analyzed utterances and verbal descriptions from experiment participants to find how people reason and talk about different levels of correlation seen in scatter-plot visualization \cite{henkin2019words}. 
Their study highlights variations between how people use correlation terms to describe a visualizations and how they actually choose to visualize the terms or phrases. 
Drawing on this and other prior work, we focus on language in the sense of probability expressions, translating them across multiple languages, and determining how people associate these expressions with icon-array visualizations (and vice versa).

\subsection{Studies Across Languages and Culture in HCI} 

Studies in human-computer interaction have investigated the impact of languages and culture on interface design and user behavior. 
In an online experiment, Baughan \etal found differences in how Japanese and American participants navigate websites, leading to concrete guidelines in which information might be presented across cultures \cite{baughan2021cross}. 
Similarly, Evers showed that peoples' understanding of a graphical interface can be influenced by their cultural experience and language, with implications for interface metaphor design and interpretation \cite{evers1998crosscultural}. 
Examining the transferability of primarily Western models of design in African contexts, Winschiers and Bidwell conduct information design activities with indigenous populations in South Africa and Namibia. Their findings surface Afro-centric paradigms which can shape interface design, with themes including cultural values such as interconnectedness, spirituality, and language used through oral and performed communication \cite{winschiers2013toward}. 
More closely related to visualization, Gibson \etal analyzed the World Color Survey in 110 languages and show that the number of color names are related to how often colors are used within a given culture. They also noted an effect of industrialization, where color becomes an essential part of the identification of objects, impacting how well people identify and name certain colors \cite{gibson2017color}. 



While it can be argued that computer interfaces are more widely distributed throughout languages and cultures than data visualizations, visualization appears to be on the rise as well.
These findings in human-computer interaction suggest that the visualization community could be doing more to question its assumptions about the universality of approaches and guidelines, particularly as data becomes more global.
We aim to take another step towards this goal by designing a study--- similar in spirit to the internationalized HCI-focused studies of LabintheWild \cite{reinecke2015labinthewild}--- to examine probability expressions in relation to icon-array visualizations, across multiple languages.


\begin{table*}[tb]
\caption{\revision{List of all expressions used in the studies and their translations. \href{https://osf.io/g5d4r/?view_only=859b329ad27847a69c8641e019ab76cf}{The supplemental materials} include this table augmented with results from the experiments, including mean assigned values and bootstrapped 95\% CIs for the \textit{Expression-to-Vis} experiment, and the ratio of usage of the expressions for each icon-array for the \textit{Vis-to-Expression} experiment.}}
\label{tab:list-probability-expressions}
\scriptsize%
\centering%
\begin{tabu}{c*{4}{l}*{1}{r}}
  \toprule 
    & English &   French &  German &  Mandarin &  Arabic  \\
  \midrule \hline
1 & plausible       & \fr{plausible}        & \de{plausibel}        & \zh{貌似可信} & \ar{معقول} \\ \hline
2 & almost certain  & \fr{presque certain}  & \de{ziemlich sicher}  & \zh{几乎确定} & \ar{شبه مؤكد} \\ \hline
3 & highly likely   & \fr{fort probable}    & \de{sehr wahrscheinlich}  & \zh{极有可能} & \ar{من المرجّح جدّا} \\ \hline
4 & very good chance& \fr{de très grandes chances} & \de{sehr gute Chance} & \zh{很有可能} & \ar{احتمال كبير} \\ \hline
5 & probable        & \fr{probable}         & \de{wahrscheinlich}   & \zh{可能} & \ar{محتمل}    \\ \hline
6 & likely          & \fr{possible}         & \de{möglich}  & \zh{或许} & \ar{مرجح}     \\ \hline
7 & probably        & \fr{probablement}     & \de{vermutlich}       & \zh{也许} & \ar{ من المحتمل}  \\ \hline
8 & chances better than even & \fr{plus d'une chance sur deux} & \de{überdurchschnittliche Chancen} & \zh{超过一半概率} & \ar{أكثر من متساوي} \\ \hline
9 & chances about even & \fr{chances à peu près égales} & \de{ungefähr gleiche Chancen}   & \zh{大约一半} & \ar{شبه متساوي} \\ \hline
10 & chances less than even & \fr{moins d'une chances sur deux} & \de{unterdurchschnittliche Chancen} & \zh{不到一半} & \ar{أقل من متساوي} \\ \hline
11 & probably not & \fr{probablement pas} & \de{wahrscheinlich nicht} & \zh{可能不会} & \multirow{2}{*}{
\ar{من غير المحتمل}}\\\cline{1-5}
12 & improbable  & \fr{improbable} & \multirow{2}{*}{\de{unwahrscheinlich}} & \zh{不太可能} & \\ \cline{1-3} \cline{5-6} 
13 & unlikely                 & \fr{invraisemblable}    &    & \zh{未必} & 
\ar{من غير المرجح} \\ \hline
14 & little chance            & \fr{peu de chance}    & \de{kleine Chance}  & \zh{没什么几率} & \ar{فرصة ضئيلة}\\ \hline
15 & almost no chance         & \fr{presque aucune chance}    & \de{fast chancenlos}   & \zh{几乎没概率} & \ar{تقريبا لا توجد فرصة} \\ \hline
16 & highly unlikely          & \fr{très peu probable}    & \de{sehr unwahrscheinlich}   & \zh{极不可能} & \ar{من المستبعد جدا} \\ \hline
17 & chances are slight       & \fr{les chances sont faibles}    & \de{die Chancen sind gering}   & \zh{机会渺茫} & \ar{فرص ضعيفة} \\ \hline
18 & implausible & \fr{peu plausible}   & \de{nicht plausibel}   &  \zh{难以置信 } & \ar{غير المعقول}
\\ 
  \bottomrule
  \end{tabu}%
\end{table*}

\section{Methodology}
 
Designing a cross-language visualization study requires addressing several challenges, primarily centered around translation, but also typical concerns such as participant scenarios/prompts, visual encoding design, and interaction.
We begin with a baseline methodology, extended from probability expression studies including Kent \cite{kent1964words}, Renooij and Witteman \cite{renooij1999talking}, and Henkin and Turkay's study in the visualization community \cite{henkin2019words}.
These inform two experiments described here that investigate (1) how people visually represent a given expression through icon-arrays, across multiple languages and (2) how people across languages choose expressions to describe a particular icon-array. 
\autoref{fig:methodology} shows an overview of our experiment methodology.




\begin{figure}[h]
    \centering
    \includegraphics[width=0.49\textwidth]{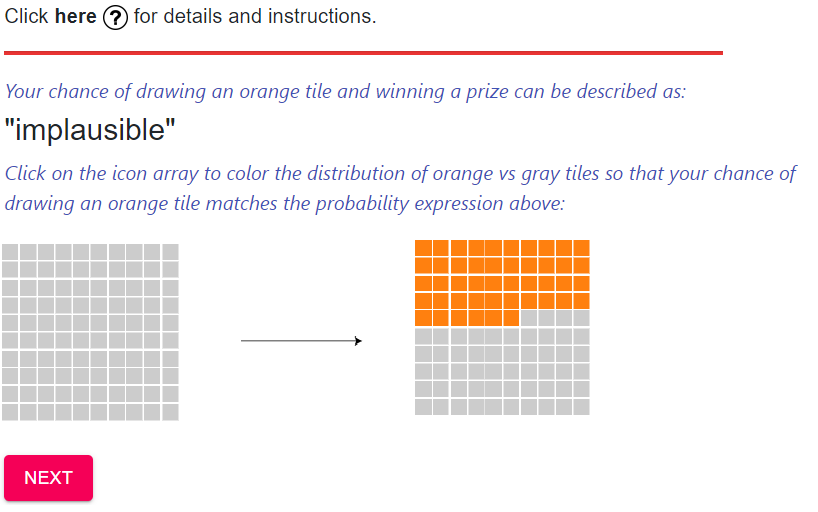} 
    \caption{A screenshot of the instructions for experiment 1 \emph{Expression-to-Vis} in English. 
    The orange-colored icon array indicates a sample answer by the participant, which we evaluate numerically as $46\%$. Participants can access the instructions at anytime during the experiment.}
    \label{fig:w2v-instructions}
\end{figure}

\subsection{Participant Prompts/Context}
In general, studies targeting probabilistic reasoning and uncertainty give participants a specific context that defines the nature of the task. 
Such framings are known to impact peoples' behavior \cite{Visschers2009}. 
Visualization studies have explored a range of scenarios, from the relatively neutral ``when is my bus coming?'' \cite{kay2016whenish} to the charged ``what is the chance that someone has cancer?'' \cite{ottley2016improving}.
Because studies have found that language and culture can impact how people perceive risk \cite{doupnik2003interpretation}, we adapt a neutral context of a game which we introduce at the beginning of each experiment as follows: 

\begin{displayquote}
You are participating in a game which consists of drawing a tile from a set at random. 
Some of the tiles are orange, and some are gray. The game has two possible outcomes:
\newline
\hspace{0.5cm}- You draw an orange tile and you win a prize \newline
\hspace{0.5cm}- You draw a gray tile and you do not win a prize 
\end{displayquote}

\subsection{Selecting and Translating Probability Expressions}

In studies targeting language and statistics, various ranges of probability expressions have been used.
For example, Kent began with five and expanded to 16 \cite{kent1964words}, while Renooij and Witteman use the seven expressions most suggested by the study participants \cite{renooij1999talking}. 
Methods for collecting expressions include scanning prior literature, eliciting expressions from study participants \cite{renooij1999talking}, and borrowing from specific documents \cite{doupnik2004impact, budescu2014interpretation}. 
\revision{While studies have examined aspects of estimative probability expressions in other languages, such as Dutch \cite{renooij1999talking} and German \cite{doupnik2003interpretation}, we did not identify comprehensive expression lists for all our target languages.}
We began with Kent's original list of 16 expressions, as they have been adapted across several studies, (\eg \cite{reagan1989quantitative, bocklisch2013doyoumean, mosteller1990quantifying, doupnik2004impact}).
Through pilot studies, we identified two ambiguities that impacted the symmetry of the list.
We add ``chances less than even” a complement to ``chances better than even” and ``implausible'' to match ``plausible''.
As a result, for this study, we use a list of 18 expressions, and provide options for participants to specify their own if none fit what they would prefer to choose during the \emph{Vis-to-Expression} experiment.

Our goal was to conduct the study in \en{English}, \fr{French}, \de{German}, \zh{Mandarin}, and \arcolor{Arabic}. 
\revision{For each language, we recruited three independent translators who were native speakers (US, France, Germany, China, Saudi Arabia, and Tunisia), and also fluent in English.}
\revision{To mitigate the potential for literal (word-by-word) translation, we reminded translators to consider the study scenario during translation, and to provide terms they would use in their native language within the given context.}

To measure agreement, we use inter-rater reliability metrics (\eg \cite{mcdonald2019reliability}) and calculate the Fleiss’ Kappa $\kappa$ values \cite{fleiss1971measuring} for the translations.
While $\kappa_{French}=0.55$, and $\kappa_{German}=0.585$ are similar, $\kappa_{Mandarin}=0.187$ and $\kappa_{Arabic}=0.341$ are lower in agreement. 
In terms of counts, the number of expressions for which all 3 translators disagree include $disag_{french} = 3$, $disag_{german}=1$, $disag_{mandarin}=8$, and $disag_{arabic}=4$. 
The numbers of expressions for which all translators agreed are $ag_{french}=8$, $ag_{german}=8$, $ag_{mandarin}=1$, and $ag_{arabic}=3$.
  
Given the high levels of disagreement in Mandarin (\ie only 9/18 expressions had two people agreeing), we engaged native speakers for possible explanations. 
One potential reason that arose from this discussion is that a group of expressions in the source language (English) can map to a group of expressions in Mandarin in an interchangeable way. 
For instance, \{“probable”, “likely”\} and \{“improbable”, “unlikely”, “probably not”\} were translated to Mandarin as \{\zh{可能, 很可能, 大概,也许}\} and \{\zh{不太可能, 可能不会, 未必}\}. 
In such cases, personal preferences might play a role in word/phrase selection.

Next we resolve disagreements in the translation.
In cases where two agree, we take the majority as the final expression. 
When all translators disagree, we use a mediation procedure until agreements are reached about the expressions \cite{mcdonald2019reliability}. 
However, providing a single translation to each English expression was not always feasible. 
For example, ``unlikely” and ``improbable” were both repeatedly translated as \de{``unwahrscheinlich”} in German, and ``probably not" and ``improbable" were both translated as 
\ar{من غير المحتمل} in Arabic.
In these cases, we reduce the number of expressions in the target language, and mark them accordingly in results.
Overall, we expect that some of these differences and similarities in languages will be reflected in the experiment results. 
\autoref{tab:list-probability-expressions} shows the final list of probability expressions used in this study with their translations in French, German, Arabic, and Mandarin.

\subsection{Visual Encodings for the Icon-Arrays} 


Visualization studies have suggested that the type and arrangement of an icon-array can impact reader perception and engagement with the underlying data \cite{zikmundfisher2014blocks, schapira2001frequency}. 
To align with the scenario described in our study, we use a 10 x 10 grid of square icons, a typical ratio in the literature for problems with a population of 100 items (used in \eg \cite{ottley2016improving} and \cite{xiong2022investigating}) . 
Color is used to denote icons representing different outcomes in the event of interest: \textit{Drawing an orange square and winning a prize}, and icons were arranged consecutively.


\subsection{Participants and Procedure}
Study participants were recruited using the online platform \href{https://www.prolific.co/}{Prolific}. 
Participants must have the target language of the experiment defined as their native language in their Prolific profile, and could only take the experiment once. 
In each study, Vis-to-Expression and Expression-to-Vis were assigned in random order, where half of the participants see Experiment 1 first while the other half see Experiment 2 first. 
50 participants were recruited for each version of the study, making a total of 250 participants for all languages. \revision{Participants were paid \$1.90 based on pilot studies estimating a 10-15 minute completion time, and following Prolific platform guidelines. Pay was fixed, \ie it was not impacted by participant choices or performance in the experiment.}


\section{Experiment 1: Expression-to-Vis, from probability expressions to icon-arrays}

In this experiment, we aim to understand how people visualize probability expressions through icon arrays, and how that varies across \en{English}, \fr{French}, \de{German}, \zh{Mandarin}, and \arcolor{Arabic}.
Unlike existing studies about numerical estimation of probability expressions, we ask participants to represent their estimations graphically.

\subsection{Procedure}

Participants see an initial icon array with only gray icons, along with a probability expression describing their chance of \emph{picking an orange tile and winning a prize}. They are asked to click or click-and-drag to draw the proportion of orange icons that matches the given probability expression. \autoref{fig:w2v-instructions} shows an example question in Experiment 1. 
The Arabic, and German versions have 17 questions, whereas the English, French, and Mandarin version have 18 questions. 
For each question, we collect the data format highlighted in the sample data in \autoref{fig:methodology} (a).

\subsection{Results}
\begin{figure}
    \centering
    \includegraphics[width=0.9\columnwidth]{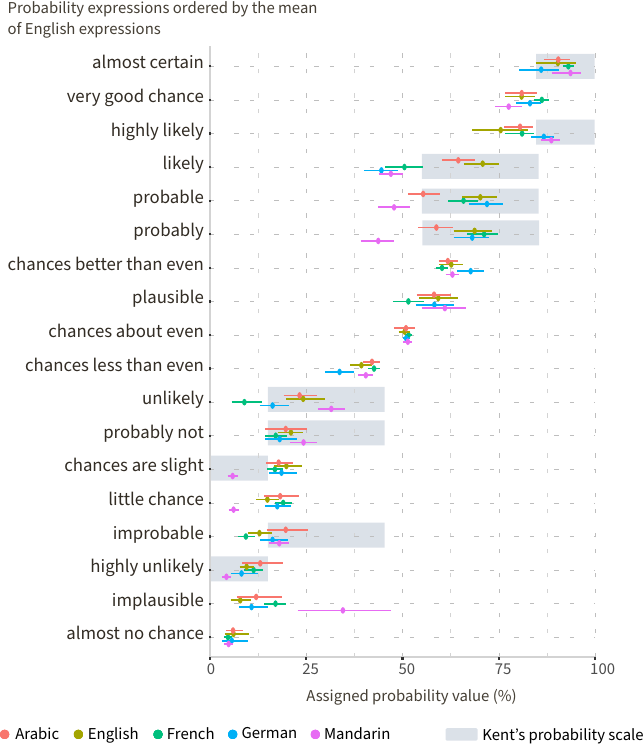}
    \caption{Bootstrapped 95\% confidence intervals of mean responses in \emph{Expression-to-Vis}. The range of responses across the five languages are tighter for expressions indicating high and low values. Shaded areas indicate the scale range of probabilities proposed by Kent \cite{kent1964words}.}
    \label{fig:w2v-95ci-kent}
\end{figure}

\begin{figure*}[th]
    \centering
    \includegraphics[width=0.94\textwidth]{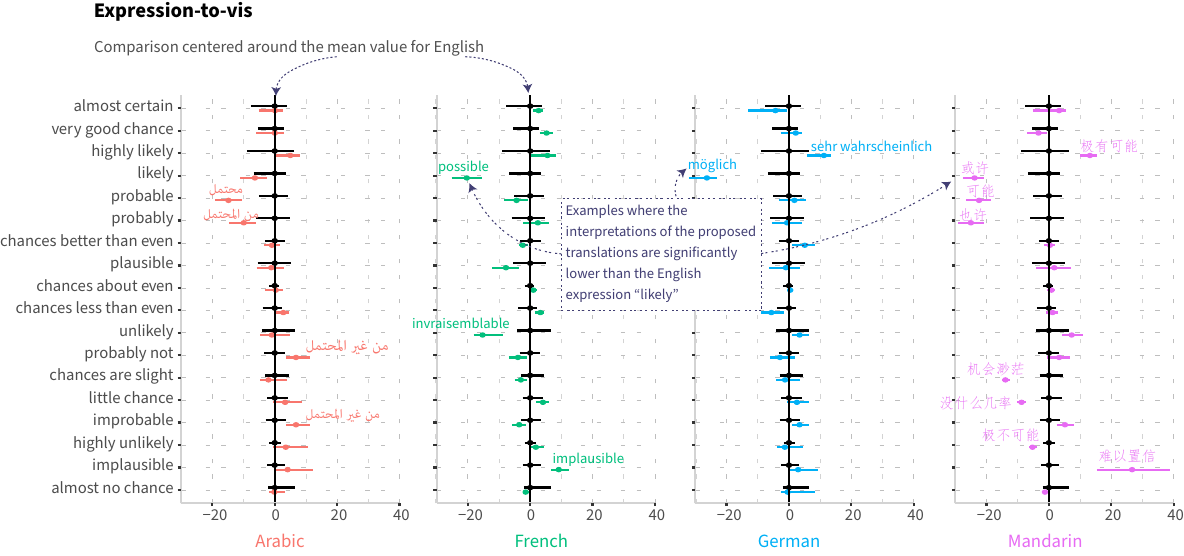}
    \caption{95\% confidence intervals for differences in mean of English expressions and their translations. Values are centered on the mean value in English. Intervals on the left indicate that the mean value for an expression is lower than its associate in English. We label translations that are significantly misaligned with the English expressions. In German, entries for ``unlikely'' and ``improbable'' are duplicated (they repeatedly translated as the same German \de{unwahrscheinlich}); similarly in Arabic , entries for ``improbable'' and ``probably not'' are duplicated (they are translated in the same Arabic expression \ar{من غير المحتمل}).
    }
    \label{fig:w2v-95ci-all-weps-centered}
\end{figure*}

In total, we collected 4,400 answers from participants across \en{English}, \fr{French}, \de{German}, \zh{Mandarin}, and \arcolor{Arabic}. 
Because there were two instances where translators agreed about a 2-to-1 mapping from an English expression to the target languages, we duplicate these confidence intervals and perform statistical comparisons separately for each. 
These include the entries for \de{unwahrscheinlich} for the English expressions ``unlikely'' and ``improbable'', and \ar{من غير المحتمل} for the English expressions ``probably not'' and ``improbable''.

\label{sec:w2v-analysis-translation}

\autoref{fig:w2v-95ci-kent} shows bootstrapped 95\% confidence intervals of the means for all expressions across the five languages. 
There appears to be general alignment with Kent's suggested ranges, though some such as highly likely and improbable deviate somewhat. 
This may be due to sample differences, \ie Kent studied 23 intelligence analysts in the 1960s.
More generally, we notice several differences for a given expression across languages.
For example, visualizations drawn for expressions aligning with ``likely'', ``probable'' and ``probably'' in French, German, Mandarin and Arabic deviate lower than English, in some cases below the 50\% mark. 

While we generally align our analyses with recommendations in the VIS and HCI communities to move beyond dichotomous statistics\cite{dragicevic2016fair}, we provide statistical comparisons here to go along with analysis shown in \autoref{fig:w2v-95ci-all-weps-centered}.
Our aim is to identify expressions that are substantially above or below the associated English translations.
While comparisons between other languages are possible, we focus on English since the expressions were originally translated from English.

Analyzing the between-language variance of participant-drawn visualizations with a one-way ANOVA, we find that only five expressions do not show at least one significant difference across the five languages. 
These stable expressions include 
``plausible"
 (and its translations: ``\fr{plausible}'', ``\de{plausibel}'', ``\ar{معقول}'', ``\zh{貌似可信}''), 
``almost certain''
 (\fr{presque certain}, \de{ziemlich sicher},\ar{شبه مؤكد}, \zh{几乎确定} ),
``chances about even'' 
 (\fr{chances à peu près égales}, \de{ungefähr gleiche Chancen}, \ar{شبه متساوي}, \zh{大约一半}), 
``probably not'' 
 (\fr{probablement pas}, \de{wahrscheinlich nicht}, \ar{من غير المحتمل}, \zh{可能不会}),
and ``almost no chance'' 
 (\fr{presque aucune chance}, \de{fast Chancenlos}, \ar{تقريبا لا توجد فرصة}, \zh{几乎没概率}).


To further analyze differences between languages, we use Tukey posthoc tests to identify pairs where the expressions significantly differ from English. \revision{All reported confidence intervals are bootstrapped 95\% CIs.}
In French, we find two deviating expressions:
    \fr{possible} (mean: -20.32, 95\% CI: [-29.23, -11.40], p.adj = 1.69E-8, English: likely), and \fr{invraisemblable} (mean: -15.26, 95\% CI: [-23.80, -6.75], p.adj = 1.65E-5, English: unlikely). 
    
Two expressions in Arabic also deviate from the English counterpart: 
    \ar{محتمل}  (mean: 14.84, 95\% CI: [6.24, 23.44], p.adj = 3.52E-5, English: probable ), and
    \ar{من المحتمل} (mean: 9.92, 95\% CI: [0.83, 19], p.adj = 0.0246, English: probably).
    
For German, we find three deviating expressions:
    \de{sehr wahrscheinlich} (mean: 11.2, 95\% CI: [2.2, 20.19], p.adj = 0.0064, English: highly likely), 
    \de{m\"{o}glich} (mean: -26.3, 95\% CI: [-35.22,-17.38], p.adj = 3.03E-13, English: likely), and 
    \de{unterdurchschnittliche Chancen} (mean: -5.66, 95\% CI: [-10.88, -0.44], p.adj = 0.026, English: chances less than even)
    
Mandarin has the highest number (seven) expressions that differ from English: 
    \zh{极有可能} (mean: 13.14, 95\% CI: [4.15, 22.13], p.adj = 0.00075, English: highly likely), 
    \zh{可能} (mean: -22.4, 95\% CI: [-31, -13.8], p.adj = 9.5E-11, English: probable),
    \zh{或许} (mean: -23.84, 95\% CI: [-32.76, -15], p.adj = 3.02E-11, English: likely), 
    \zh{也许} (mean: -25.02, 95\% CI: [-34.10, -15.93], p.adj = 7.76E-12, English: probably), 
    \zh{没什么几率} (mean: -8.78, 95\% CI: [-15.13, -2.42], p.adj = 0.0017, English: little chance),
    \zh{机会渺茫} (mean: -13.94, 95\% CI: [-19.9, -7.98], p.adj = 6.44E-9, English: chances are slight), and
    \zh{难以置信} (mean: 26.66, 95\% CI: [13.9, 39.42], p.adj = 2.76E-7, English: implausible). 
    
These results show multiple instances where participants in a particular language consistently draw icon-arrays that align with different probability ranges than the associated English expression, both above and below.
We will discuss possible reasons behind these differences, including implications for visualization, in \autoref{sec:discussion}.
Interestingly, translations for the duplicated entries both in German and Arabic did not significantly differ from the original English expression. This suggests that the suggested translation in German and Arabic do align with both expressions in English.


\begin{figure*}[h]
    \centering
        \includegraphics[width=\textwidth]{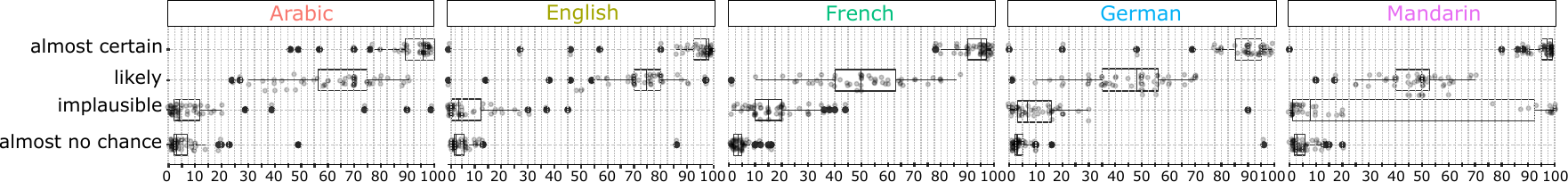}
    \caption{Range plots of example expressions and values drawn on the icon-arrays (Expression-to-Vis). There seems to be rounding behavior around multiples of 0 and 5. Value ranges for some expressions also vary across translations (\eg ``likely'' differs between English and other languages).} 
    \label{fig:w2v-boxplot-sample}
\end{figure*}

Other patterns in the analysis reflect possible drawing affordances in the icon-array that persist across languages.
Most responses end in 0 or 5, similar to rounding behavior in graphical perception studies \cite{talbot2014four}.
These include 51\% of total answers for English, 54\% for French, 52\% for German, 45\% for Arabic, 48\% for Mandarin. 
This pattern also aligns with previous results where people tend provide numerical estimations that are multiples of 10 for verbal probabilities \cite{leclercq2016jensuis, willems2020variability} (see \autoref{fig:w2v-boxplot-sample})

We can also view each base expression from the perspective of its range across languages, see \autoref{fig:w2v-boxplot-sample}.
For example, although there is no particular pattern across the five languages for the expressions, the Mandarin translation of ``implausible" shows the largest IQR with 91\%, \revision{as well as a bimodal distribution of responses}. 
Expressions with narrow ranges suggest that, across associated expressions in other languages, participants will draw similar ranges in icon-arrays. 
Another pattern is that expressions near extreme low/high and center values appear to have smaller interquartile ranges. 
Expressions at the extremes are consistently evaluated, while mid-range expressions (but not central) may convey less precise estimates of probability.

\begin{figure}[h]
    \centering
    \includegraphics[width=0.5\textwidth]{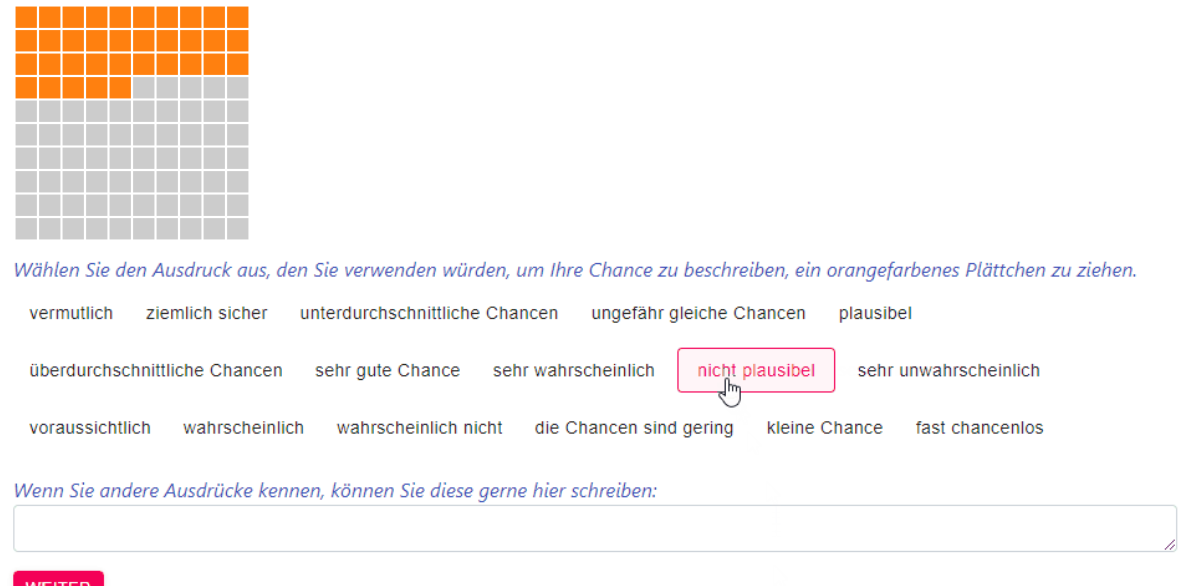}
    \caption{A screenshot of the instructions for experiment 2 \emph{Vis-to-Expression} in German. Participants were also given the option to input their own expression to describe the icon-array.}
    \label{fig:v2w-instructions}
\end{figure}

\section{Experiment 2: Vis-to-Expression, From icon-arrays to probability expressions}

This experiment is essentially an inversion of Experiment 1. Here, we aim to understand how people choose probability expressions to describe a given icon-array visualization. 

\subsection{Procedure}
Using the same neutral scenario, participants are given an icon-array of a specific value, along with a list of probability expressions (see \autoref{fig:v2w-instructions}. 
Participants are asked to select the expression that they believe best describes the icon-array shown. 
Expression lists consist of 18 expressions in \en{English}, \fr{French}, and \zh{Mandarin}, 17 expressions in \de{German}, and \arcolor{Arabic}.
Participants are also encourage to provide their own answer, if desired.
To cover the probability space, we encode 19 values between 5\% to 95\% with a step of 5\%.
For each trial, we collect the pair \{icon-array value, selected probability expression\}. 

\subsection{Results}
\begin{figure*}[th]
    \centering
    \includegraphics[width=0.9\textwidth]{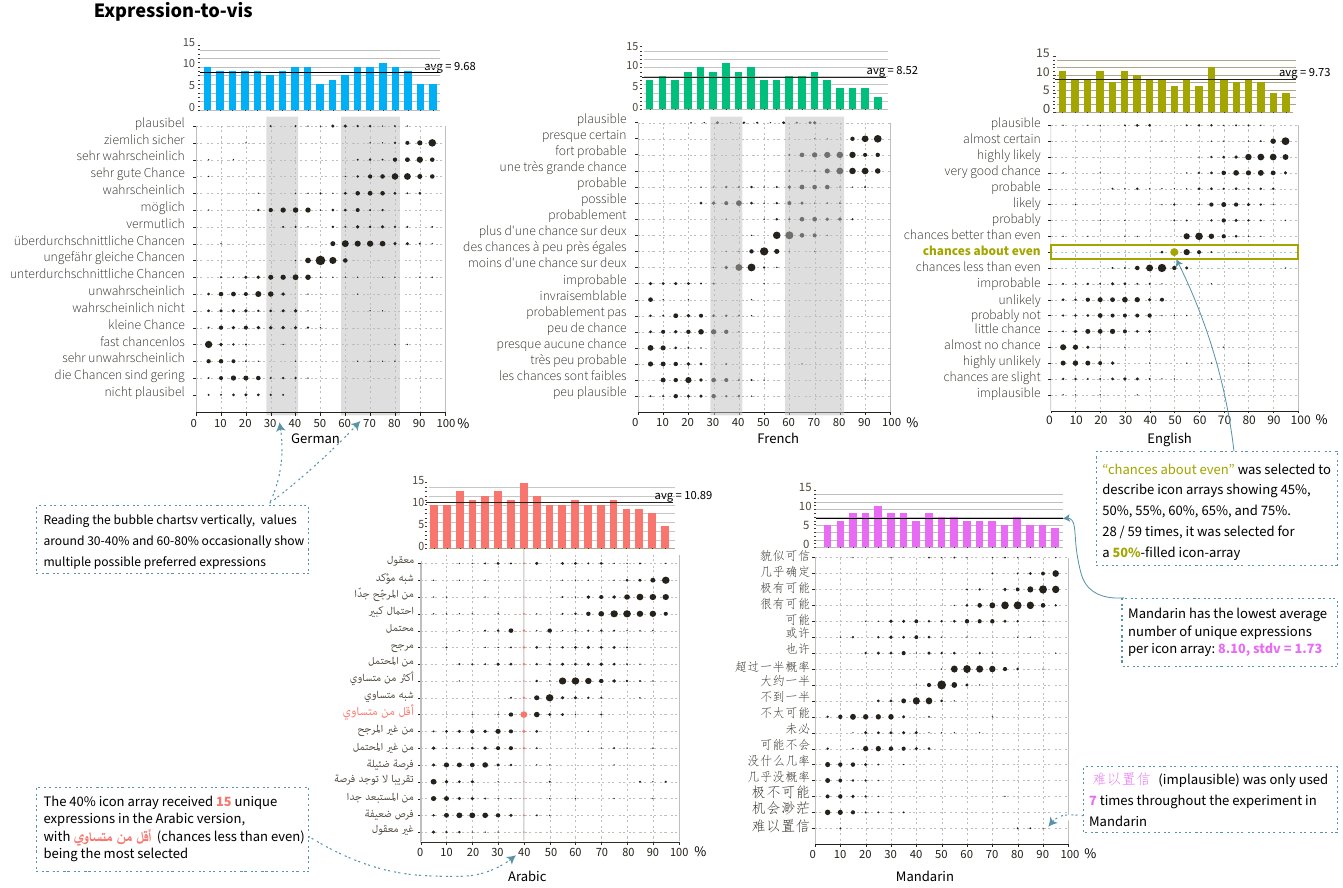}
    \caption{Results from experiment 2 \emph{Expression-to-Vis}. Barcharts represent the number of unique expressions participants selected for a given icon-array. Bubbles indicate count of each probability expression and value pair. Results show similarities and differences across languages.}
    \label{fig:v2w-scatterplot}
\end{figure*}

\revision{In total, we collect 4,750 icon-array to expression pairs from participants across the five languages (50 participants per language). Icon arrays across steps of 5\% are described fifty times by participants, either by selecting an expression from the suggested list, or by typing their answers. Similar to Expression-to-Vis, we duplicate entries for expressions that map to two expressions in English. These include \ar{من غير المحتمل} in Arabic and \de{unwahrscheinlich} in German. Results for Vis-to-Expression are shown in \autoref{fig:v2w-scatterplot}.}

The upper bar graph in \autoref{fig:v2w-scatterplot} shows the count of unique expressions for each icon-array value. 
Across languages, there appears to be consistency in that higher values (90\%, and 95\%) are described using fewer unique expressions. 
This may reflect a more consistent expressiveness of languages of probability expressions in higher ranges. 
However, given the average counts across all value possibilities, it is clear that few, if any perfect matches of probability expressions and visualization exists.

Arabic has the highest average number of unique expressions. 
On average, 10.89 expressions (stdv = 2) were used to describe each probability value. 
For example, for icon-arrays of 40\%, participants in Arabic chose 15 out of 17 possible expressions.
10 of these expressions were selected at least twice.
Looking back at the Expression-to-Vis results in \autoref{fig:w2v-95ci-kent}, there appears to be a gap above 25\% and below 45\%, although \ar{أقل من متساوي} or ``chances less than even'' did end up being the most selected expression for this value (19 times).

In contrast, Mandarin has the lowest average count of unique expressions per icon array 8.10 (stdv = 1.73) than the other versions of the experiment.
Implausible is a notable exception, which was rarely used and had a wide range associated with it.
To a lesser degree, German, French, and English show the central values around 50\% with a low number of unique expressions (below their averages), while mid-range values above and below 50\% show more variance. 

\autoref{fig:v2w-scatterplot} also shows the intersection of icon-array values and expression counts. 
The size of the circles indicate the ratio at which an expressions has been used to describe a given value. 

Looking vertically, there appears to be variance in how icon-arrays depicting particular values are described, with some expressions preferred over other. 
For example, values around 30-40\% and 60-80\% occasionally show multiple possible preferred expressions.
Across all values and expressions, larger circles appear to exist for central and high probability values, reflecting some of the consistency seen in the first experiment. 
People across languages seem to have clear preferences for translations of ``chances about even" for an icon-array at 50\%. 
Other patterns show variance and disagreement.
The translations for ``probable", ``likely" appear to be used to specify a large range of probability values in a high range.
Notably, however, similar patterns do not appear to exist across languages for wide ranges of low probability values.
For instance, while ``implausible'' translations potentially fits this low-and-wide range role is German, French, and Arabic, it is scarcely chosen at all in English and Mandarin.

During the experiment, participants had the option to suggest additional probability expressions whenever needed. 
Participants provided their own expressions \arcolor{55} times  for Arabic, \en{99} for English, \fr{145} for French, \de{65} for German, and \zh{14} for Mandarin. 
Example trends in these include people using the listed expressions in a full sentences (\eg \fr{ Il est \emph{improbable} de gagner} (``it is improbable to win")), or with varying qualifiers (\eg \ar{فرصة عالية, فرصة عالية جدا}
 (``high chance, very high chance")). 
Among the new expressions and phrases that were suggested, we notice some referring directly to the proportion shown in the icon array arrangement (\eg \de{ Etwa jeder Vierte gewinnt} for a 25\% icon array). 
We provide these data in the supplement, as an extended analysis of written answers by participants may give an opportunity to explore additional language or cultural factors that people refer to when making judgments about icon-array visualizations. 


\section{\revision{Comparing Expression Usage Across Experiments}}\label{sec:comparing-two-exp}

\begin{figure}[ht]
    \includegraphics[width=0.45\textwidth, angle = 0]{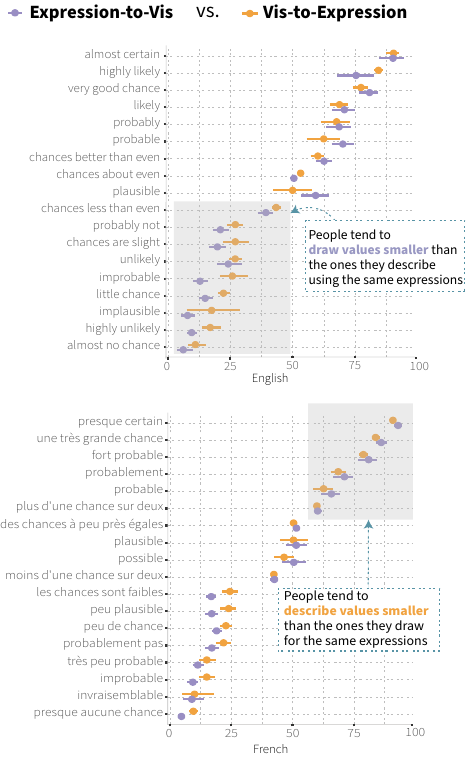}
    \caption{Comparing the two experiments, results show differences in elicitation methodology that hold across languages. People appear to draw visualizations with more extreme values when given an expression. But when given a visualization of similar value, people tend to choose different expressions. Charts for all five languages and for individual participants are available in \href{https://osf.io/g5d4r/?view_only=859b329ad27847a69c8641e019ab76cf}{supplementary material}.} 
    \label{fig:compare-exp}
\end{figure}

Given the within-subjects design of both experiments, it is possible to make comparisons across them.
In experiment 1 \emph{Expression-to-Vis}, people were given each expression and drew a specific icon-array design (stored as a percentage value). 
Similarly, In experiment 2 \emph{Vis-to-Expression}, icon-arrays of particular values were given and participants chose expressions. 
Shown in \autoref{fig:compare-exp}, one observation between the two experiments is that people tend to draw extreme values for given expressions, but when people are given icon-arrays depicting similar values, they choose different expressions. 
A concrete example is that when someone receives an expression \emph{``almost no chance"}, they typically draw an icon array with very few colored squares. 
However, when showed with an icon-array with more colored squares than are typically drawn for this expression, people still describe this icon-array as ``almost no chance''. 

With these values for each participant, we can also explore how consistent people are in assigning expressions to visualizations versus visualizations to expressions. 
While several distance metrics are possible, in an exploratory analysis we define a straightforward distance measure for each probability expression as 
    \emph{Distance D = exp2vis\_val - mean(vis2exp\_val)}.
Where $exp2vis\_val$ is the value that a participant drew for an expression, and $mean(vis2exp\_val)$ is the average of all values for which participants assigned this expression.
A large distance indicates that a participant is less consistent their mapping of language to visualization, while a low distance suggests more consistency.
We find several instances where participants exhibit consistency across the two experiments, and other instances where participants differ between themselves widely. 
Example visualizations and data for these are included in the supplement.
These also motivate potential individual-level modeling efforts for future work.





\section{Discussion} \label{sec:discussion}

The results of the \emph{Expression-to-Vis} and \emph{Vis-To-Expression} experiments suggest that language plays an important role in the specification and interpretation of icon-array visualizations.
Differences are found within languages. For example, in Mandarin, there was no overlap in Kent's suggested scale range of 55\%-85\% for the expressions \emph{likely, probable}, and \emph{probably} and the visualizations participants drew when given translations of these expressions.

There are other differences across languages.
Using English as a source translation, we identified instances in every tested language that deviated significantly above or below the ranges for the associated English expression (\autoref{fig:w2v-95ci-kent}).
These and other reported findings raise questions about the interplay between language expressiveness, visualization, and translation efforts.

\subsection{Implications for Visualization Translation}
\revision{The observed differences across languages may hold implications for visualization translation efforts. Consider a case where a climate change visualization originally crafted in English, but is intended to be translated into other languages for broader distribution. 
Translation may use a similar visualization technique, but requires substantial changes to the accompanying textual elements to facilitate target audiences' understanding of the communicated data in the specific context of the visualization. 
Results here suggest that estimative probability expressions, which may be used both in visualizations themselves and in texts referencing the visualization, should be carefully translated so as to preserve the intended numerical range.
Difficulties in translation may persist across both translators and automatic translation tools, such as Google Translate. 
Future studies might build on these results and others examining the impact of text on visualization interpretation, \eg \cite{kong2018frames}, to investigate the extent to which alignments and misalignments between expressions and visualizations impact peoples' judgments.}


The translation process itself is another factor that may play a role in the observed differences. 
Although multiple native speakers were consulted and mediation procedures were used to reach agreement, it is possible that other speakers would have generated somewhat different expressions.
Even with other translators, however, there is no guarantee that the chosen expressions would have matched the ranges participants chose for the source expression.
A different resulting expression list would also not necessarily cover equal spans of the probability value ranges in the target language.

These translation challenges raise one possible application of the methodology and the results here.
It may be possible to \emph{align} expressions based on the resulting participant-driven probability ranges.
For example, while the expression \fr{plausible} in French differs significantly from the English \en{plausible}, another expression in French, \fr{plus d'une chance sur deux} (originally translating \emph{chances better than even}), does align better with this range in the observed data.
Computational methods could be designed to construct these translations for the tested languages and others.
As an initial exploration, we iteratively paired up expressions across two languages until we found pairs that do not significantly differ following a Tukey test comparison. 
The results of this approach showing multiple possible ``aligned'' translations are available in supplemental material.

Computational approaches to language-expression alignment could address challenges in cross-language statistical reporting scenarios.
The IPCC (Intergovernmental Panel on Climate Change) report, for example, specifies guidelines for its writers to use certain probability expressions for certain ranges \cite{budescu2014interpretation}. 
As medical tests and associated symptom displays also rely on an intersection of icon-arrays and language (\eg \cite{ottley2016improving, garciaretamero2009communicating, garciaRetamero2010profits}), these efforts may also aid medical or pandemic risk communication.
Our results provide a possible path towards further refining these standards, helping ensure the intended meaning of statistics and charts is communicated faithfully across languages.


\subsection{Towards Better Elicitation for Cross-Language Visualization}


Methodology was a key challenge in the study.
While we aimed to carefully adapt and extend prior studies to begin exploring the intersection of visualization and languages, possibilities emerged through the design and resulting analysis.
In the \emph{Vis-to-Expression} experiment, participants offered 86 (out of 4750) additional expressions. While some of these overlap with existing expressions in the study, it is likely that there are other expressions or phrases that the studied languages and associated cultures use in talking about probabilistic events.
Finding ways to elicit these could be a challenging but rewarding effort for visualization.
For instance, a participant in the Mandarin version commented: 
\begin{displayquote}
\zh{"貌似可信”在汉语中不是一个好的表达，我作为母语者都不能完全理解你们用这个词想表达什么"} (It seems ``plausible" is not a good expression in Chinese. As a native speaker, I can't fully understand what you are trying to express with this word)
\end{displayquote}

In English, ``plausible" is common, but might there be other translations or similar expressions in other languages that fill a similar role?

One possibility is to move beyond English as a source, and instead develop in-language elicitation methodologies.
These might be graphically based, using interaction and visualization with input capability to allow participants to specify ranges and expressions.
Alternately, they may be large crowdsourced studies, following similar scenario and trial-based methods.
In either case, the goal would be to elicit a wider range of expressions and visualization descriptions from participants.
Such efforts could reveal additional ranges and expressive capabilities within languages, beyond those studied here.

Beyond the results presented here, there are other useful starting points for exploring possibilities in cross-language elicitation for visualization. 
One source would be to consider the history of large-scale color elicitation studies such as the World Color Survey \cite{lindsey2009world}.
Other language and statistics studies such as Renooij and Witteman\cite{renooij1999talking}, and the NLP-driven analysis of Henkin and Turkay \cite{henkin2019words} might inform approaches to scale.

\subsection{Limitations and Future Work}

One limitation of the current work is the restriction to five languages.
While these were chosen as an initial step in the space, and intended to cover several major language families, there are thousands of languages in the world and various language groups that could be explored.
In the context of a global pandemic, for example, it is important to support effective data-focused communication as broadly as possible.
\revision{Another limitation is the use of English as a source language. While the intended focus of this study is to highlight difficulties surrounding translation, the use of English still somewhat centers findings on WEIRD populations. Other possible study designs are within reach, however. Future work might design expression elicitation methods directly for native speakers, or using both visualization and numerical probabilities, which might reveal additional expressive features between languages. Elicitation methods might also examine the language people use when describing visualizations of probability distributions, instead of individual probabilities.}
The neutral scenario / context given to participants and sole use of icon-array visualizations are other practical limitations to explore in future work. 
Cultural differences such as risk avoidance (\eg \cite{bocklisch2013doyoumean, doupnik2003interpretation}) may be less pronounced in neutral scenarios, but become more pronounced with carefully designed contexts.
We might refer to studies targeting medical reasoning (\eg Ottley \etal \cite{ottley2016improving}) or natural disaster risk (\eg Padilla \etal \cite{padilla2020uncertainty}) for promising scenarios and visualizations to explore in future work.
\revision{Finally, to more directly target implications for visual representations, future studies might explore the extent to which \emph{different} visual representations elicit responses from native speakers, which might uncover effects of language and culture between visualizations.}

\section{Conclusion}

With the changing global landscape of data and visualization practice, it is important that the research community explores the intersection of language and visualization.
We present two experiments with the goal of understanding how people across five languages draw icon-array visualizations given probability expressions and assign probability expressions given icon-array visualizations.
Results of these experiments show several differences both across and within languages, with no clear mapping across languages, and several instances of possible ``gaps'' between expressions in a given language. 
We discuss implications of these results for ongoing efforts such as data and visualization translation, targeting areas such as climate change and pandemic communication.
Taken together, these studies and results are intended to offer a limited yet useful step in broadening the focus of the visualization community beyond traditional WEIRD populations.



\acknowledgments{
The authors wish to thank Alvitta Ottley and Melanie Bancilhon, as well as Cagatay Turkay and Rafael Henkin for their input on early versions of this work. The authors also wish to thank the translators, and the many participants on the Prolific.co platform who participated in the study.
}

\bibliographystyle{abbrv-doi}

\bibliography{template}

\begin{thebibliography}{10}

\bibitem{bancilhon2019gamble}
M.~Bancilhon, Z.~Liu, and A.~Ottley.
\newblock {Let's Gamble: Uncovering the Impact of Visualization on Risk
  Perception and Decision-Making}.
\newblock {\em Computing Research Repository (CoRR)}, abs/1910.09725, 2019.

\bibitem{barclay1977handbook}
S.~Barclay et~al.
\newblock Handbook for decisions analysis.
\newblock 1977.

\bibitem{baughan2021cross}
A.~Baughan, N.~Oliveira, T.~August, N.~Yamashita, and K.~Reinecke.
\newblock {Do Cross-Cultural Differences in Visual Attention Patterns Affect
  Search Efficiency on Websites?}
\newblock In {\em {CHI} '21: {CHI} Conference on Human Factors in Computing
  Systems, Virtual Event / Yokohama, Japan, May 8-13, 2021}, pp. 362:1--362:12.
  {ACM}, 2021. doi: {{%
10\hspace{.1pt}\discretionary{.}{%
}{.}\hspace{.4pt}1145\discretionary{/}{%
}{/}3411764\hspace{.1pt}\discretionary{.}{%
}{.}\hspace{.4pt}3445519}}


\bibitem{bocklisch2013doyoumean}
F.~Bocklisch, A.~Georg, S.~Bocklisch, and J.~Krems.
\newblock {Do You Mean What You Say? The Effect of Uncertainty Avoidance on the
  Interpretation of Probability Expressions-A Comparative Study between Spanish
  and German}.
\newblock In {\em Proceedings of the Annual Meeting of the Cognitive Science
  Society}, vol.~35. cognitivesciencesociety.org, 2013.

\bibitem{brustrenck2013communicating}
P.~G. Brust-Renck, C.~E. Royer, and V.~F. Reyna.
\newblock {Communicating Numerical Risk: Human Factors That Aid Understanding
  in Health Care}.
\newblock {\em Reviews of Human Factors and Ergonomics}, 8(1):235--276, 2013.
  doi: {{%
10\hspace{.1pt}\discretionary{.}{%
}{.}\hspace{.4pt}1177\discretionary{/}{%
}{/}1557234X13492980}}


\bibitem{budescu2014interpretation}
D.~V. Budescu, H.~H. Por, S.~B. Broomell, and M.~Smithson.
\newblock {The Interpretation of {IPCC} Probabilistic Statements around the
  World}.
\newblock {\em Nature Climate Change}, 4(6):508--512, 2014. doi: {{%
10\hspace{.1pt}\discretionary{.}{%
}{.}\hspace{.4pt}1038\discretionary{/}{%
}{/}nclimate2194}}


\bibitem{budescu1985consistency}
D.~V. Budescu and T.~S. Wallsten.
\newblock {Consistency in Interpretation of Probabilistic Phrases}.
\newblock {\em {Organizational behavior and human decision processes}},
  36(3):391--405, 1985.

\bibitem{cleveland1984graphical}
W.~S. Cleveland and R.~McGill.
\newblock {Graphical Perception: Theory, Experimentation, and Application to
  the Development of Graphical Methods}.
\newblock {\em Journal of the American statistical association},
  79(387):531--554, 1984.

\bibitem{doupnik2006influence}
T.~S. Doupnik and E.~L. Riccio.
\newblock {The Influence of Conservatism and Secrecy on the Interpretation of
  Verbal Probability Expressions in the Anglo and Latin Cultural Areas}.
\newblock {\em The International Journal of Accounting}, 41(3):237--261, 2006.

\bibitem{doupnik2003interpretation}
T.~S. Doupnik and M.~Richter.
\newblock {Interpretation of Uncertainty Expressions: a Cross-national Study}.
\newblock {\em Accounting, Organizations and Society}, 28(1):15--35, 2003. doi:
  {{%
10\hspace{.1pt}\discretionary{.}{%
}{.}\hspace{.4pt}1016\discretionary{/}{%
}{/}S0361\discretionary{%
}{-}{-}3682\discretionary{%
}{(}{(}02\discretionary{)}{%
}{)}00010\discretionary{%
}{-}{-}7}}


\bibitem{doupnik2004impact}
T.~S. Doupnik and M.~Richter.
\newblock {The Impact of Culture on the Interpretation of “In Context”
  Verbal Probability Expressions}.
\newblock {\em Journal of International Accounting Research}, 3(1):1--20, 2004.
  doi: {{%
10\hspace{.1pt}\discretionary{.}{%
}{.}\hspace{.4pt}2308\discretionary{/}{%
}{/}jiar\hspace{.1pt}\discretionary{.}{%
}{.}\hspace{.4pt}2004\hspace{.1pt}\discretionary{.}{%
}{.}\hspace{.4pt}3\hspace{.1pt}\discretionary{.}{%
}{.}\hspace{.4pt}1\hspace{.1pt}\discretionary{.}{%
}{.}\hspace{.4pt}1}}


\bibitem{dragicevic2016fair}
P.~Dragicevic.
\newblock {Fair Statistical Communication in HCI}.
\newblock In {\em Modern statistical methods for HCI}, pp. 291--330. Springer,
  2016.

\bibitem{evers1998crosscultural}
V.~Evers.
\newblock {Cross-Cultural Understanding of Metaphors in Interface Design}.
\newblock {\em Proceedings CATAC'98, Cultural Attitudes towards Technology and
  Communication}, pp. 1--11, 1998.

\bibitem{fleiss1971measuring}
J.~L. Fleiss.
\newblock {Measuring Nominal Scale Agreement among Many Raters.}
\newblock {\em Psychological bulletin}, 76(5):378, 1971.

\bibitem{fuhrman2010cross}
O.~Fuhrman and L.~Boroditsky.
\newblock {Cross-Cultural Differences in Mental Representations of Time:
  Evidence from an Implicit Nonlinguistic Task}.
\newblock {\em Cognitive science}, 34(8):1430--1451, 2010.

\bibitem{galesic2009using}
M.~Galesic, R.~Garcia-Retamero, and G.~Gigerenzer.
\newblock {Using Icon Arrays to Communicate Medical Risks: Overcoming Low
  Numeracy}.
\newblock {\em Health Psychology}, 28(2):210--216, 2009. doi: {{%
10\hspace{.1pt}\discretionary{.}{%
}{.}\hspace{.4pt}1037\discretionary{/}{%
}{/}a0014474}}


\bibitem{garciaretamero2009communicating}
R.~Garcia-Retamero and M.~Galesic.
\newblock {Communicating Treatment Risk Reduction to People with Low Numeracy
  Skills: A Cross-Cultural Comparison}.
\newblock {\em American Journal of Public Health}, 99(12):2196--2202, 2009.
  doi: {{%
10\hspace{.1pt}\discretionary{.}{%
}{.}\hspace{.4pt}2105\discretionary{/}{%
}{/}AJPH\hspace{.1pt}\discretionary{.}{%
}{.}\hspace{.4pt}2009\hspace{.1pt}\discretionary{.}{%
}{.}\hspace{.4pt}160234}}


\bibitem{garciaRetamero2010profits}
R.~Garcia-Retamero and M.~Galesic.
\newblock {Who Profits from Visual Aids: Overcoming Challenges in People's
  Understanding of Risks}.
\newblock {\em Social Science and Medicine}, 70(7):1019--1025, 2010. doi: {{%
10\hspace{.1pt}\discretionary{.}{%
}{.}\hspace{.4pt}1016\discretionary{/}{%
}{/}j\hspace{.1pt}\discretionary{.}{%
}{.}\hspace{.4pt}socscimed\hspace{.1pt}\discretionary{.}{%
}{.}\hspace{.4pt}2009\hspace{.1pt}\discretionary{.}{%
}{.}\hspace{.4pt}11\hspace{.1pt}\discretionary{.}{%
}{.}\hspace{.4pt}031}}


\bibitem{gibson2017color}
E.~Gibson, R.~Futrell, J.~Jara-Ettinger, K.~Mahowald, L.~Bergen,
  S.~Ratnasingam, M.~Gibson, S.~T. Piantadosi, and B.~R. Conway.
\newblock {Color Naming Across Languages Reflects Color Use}.
\newblock {\em Proceedings of the National Academy of Sciences},
  114(40):10785--10790, 2017.

\bibitem{gigerenzer1995improve}
G.~Gigerenzer and U.~Hoffrage.
\newblock {How to Improve Bayesian Reasoning without Instruction: Frequency
  Formats}.
\newblock {\em Psychological Review}, 102(4):684--704, 1995. doi: {{%
10\hspace{.1pt}\discretionary{.}{%
}{.}\hspace{.4pt}1037\discretionary{/}{%
}{/}0033\discretionary{%
}{-}{-}295X\hspace{.1pt}\discretionary{.}{%
}{.}\hspace{.4pt}102\hspace{.1pt}\discretionary{.}{%
}{.}\hspace{.4pt}4\hspace{.1pt}\discretionary{.}{%
}{.}\hspace{.4pt}684}}


\bibitem{henkin2019words}
R.~Henkin and C.~Turkay.
\newblock {Words of Estimative Correlation: Studying Verbalizations of
  Scatterplots}.
\newblock {\em CoRR}, abs/1911.12793, 2019.

\bibitem{kay2016whenish}
M.~Kay, T.~Kola, J.~R. Hullman, and S.~A. Munson.
\newblock {When (ish) is my Bus? User-Centered Visualizations of Uncertainty in
  Everyday, Mobile Predictive Systems}.
\newblock {\em Conference on Human Factors in Computing Systems - Proceedings},
  pp. 5092--5103, 2016. doi: {{%
10\hspace{.1pt}\discretionary{.}{%
}{.}\hspace{.4pt}1145\discretionary{/}{%
}{/}2858036\hspace{.1pt}\discretionary{.}{%
}{.}\hspace{.4pt}2858558}}


\bibitem{kent1964words}
S.~Kent.
\newblock {Words of Estimative Probability}.
\newblock {\em Journal of the American Intelligence Professional}, 8(4):49--65,
  1964.

\bibitem{kim2019color}
Y.~Kim, K.~Thayer, G.~S. Gorsky, and J.~Heer.
\newblock {Color Names Across Languages: Salient Colors and Term Translation in
  Multilingual Color Naming Models.}
\newblock In {\em EuroVis (Short Papers)}, pp. 31--35, 2019.

\bibitem{kong2018frames}
H.-K. Kong, Z.~Liu, and K.~Karahalios.
\newblock Frames and slants in titles of visualizations on controversial
  topics.
\newblock In {\em Proceedings of the 2018 CHI conference on human factors in
  computing systems}, pp. 1--12, 2018.

\bibitem{kreuzmair2016high}
C.~Kreuzmair, M.~Siegrist, and C.~Keller.
\newblock {High Numerates Count Icons and Low Numerates Process Large Areas in
  Pictographs: Results of an Eye-Tracking Study}.
\newblock {\em Risk Analysis}, 36(8):1599--1614, 2016. doi: {{%
10\hspace{.1pt}\discretionary{.}{%
}{.}\hspace{.4pt}1111\discretionary{/}{%
}{/}risa\hspace{.1pt}\discretionary{.}{%
}{.}\hspace{.4pt}12531}}


\bibitem{leclercq2016jensuis}
D.~Leclercq.
\newblock {J'en Suis Aussi S{\^{u}}r que Vous, Mais pas avec le M{\^{e}}me
  Pourcentage de Chances, que ce Soit Hors Contexte ou En Contexte}.
\newblock {\em Journal international de Recherche en Education et Formation
  Evaluer. Journal international de Recherche en Education et Formation},
  2(21):89--125, 2016.

\bibitem{lindsey2009world}
D.~T. Lindsey and A.~M. Brown.
\newblock {World Color Survey Color Naming Reveals Universal Motifs and their
  Within-Language Diversity}.
\newblock {\em Proceedings of the National Academy of Sciences of the United
  States of America}, 106(47):19785--19790, 2009. doi: {{%
10\hspace{.1pt}\discretionary{.}{%
}{.}\hspace{.4pt}1073\discretionary{/}{%
}{/}pnas\hspace{.1pt}\discretionary{.}{%
}{.}\hspace{.4pt}0910981106}}


\bibitem{mcdonald2019reliability}
N.~McDonald, S.~Schoenebeck, and A.~Forte.
\newblock {Reliability and Inter-Rater Reliability in Qualitative Research:
  Norms and Guidelines for CSCW and HCI Practice}.
\newblock {\em Proceedings of the ACM on Human-Computer Interaction}, 3(CSCW),
  2019. doi: {{%
10\hspace{.1pt}\discretionary{.}{%
}{.}\hspace{.4pt}1145\discretionary{/}{%
}{/}3359174}}


\bibitem{mosteller1990quantifying}
F.~Mosteller and C.~Youtz.
\newblock {Quantifying Probabilistic Expressions}.
\newblock {\em Statistical Science}, 5(1):2--12, 1990. doi: {{%
10\hspace{.1pt}\discretionary{.}{%
}{.}\hspace{.4pt}1214\discretionary{/}{%
}{/}ss\discretionary{/}{%
}{/}1177012242}}


\bibitem{okan2015improving}
Y.~Okan, R.~Garcia-Retamero, E.~T. Cokely, and A.~Maldonado.
\newblock {Improving Risk Understanding across Ability Levels: Encouraging
  Active Processing with Dynamic Icon Arrays}.
\newblock {\em Journal of Experimental Psychology: Applied}, 21(2):178--194,
  2015. doi: {{%
10\hspace{.1pt}\discretionary{.}{%
}{.}\hspace{.4pt}1037\discretionary{/}{%
}{/}xap0000045}}


\bibitem{ottley2016improving}
A.~Ottley, E.~M. Peck, L.~T. Harrison, D.~Afergan, C.~Ziemkiewicz, H.~A.
  Taylor, P.~K. Han, and R.~Chang.
\newblock {Improving Bayesian Reasoning: The Effects of Phrasing,
  Visualization, and Spatial Ability}.
\newblock {\em IEEE Transactions on Visualization and Computer Graphics},
  22(1):529--538, 2016. doi: {{%
10\hspace{.1pt}\discretionary{.}{%
}{.}\hspace{.4pt}1109\discretionary{/}{%
}{/}TVCG\hspace{.1pt}\discretionary{.}{%
}{.}\hspace{.4pt}2015\hspace{.1pt}\discretionary{.}{%
}{.}\hspace{.4pt}2467758}}


\bibitem{padilla2021review}
L.~Padilla, S.~C. Castrob, and H.~Hosseinpoura.
\newblock {A Review of Uncertainty Visualization Errors: Working Memory as an
  Explanatory Theory}.
\newblock {\em The Psychology of Learning and Motivation}, p. 275, 2021.

\bibitem{padilla2020uncertainty}
L.~Padilla, M.~Kay, and J.~Hullman.
\newblock {Uncertainty Visualization}.
\newblock {\em Handbook of Computational Statistics and Data Science}, 2020.

\bibitem{peck2019data}
E.~M. Peck, S.~E. Ayuso, and O.~El-Etr.
\newblock {Data is Personal: Attitudes and Perceptions of Data Visualization in
  Rural Pennsylvania}.
\newblock {\em arXiv}, pp. 1--12, 2019.

\bibitem{reagan1989quantitative}
R.~T. Reagan, F.~Mosteller, and C.~Youtz.
\newblock {Quantitative Meanings of Verbal Probability Expressions}.
\newblock {\em Journal of Applied Psychology}, 74(3):433--442, 1989. doi: {{%
10\hspace{.1pt}\discretionary{.}{%
}{.}\hspace{.4pt}1037\discretionary{/}{%
}{/}0021\discretionary{%
}{-}{-}9010\hspace{.1pt}\discretionary{.}{%
}{.}\hspace{.4pt}74\hspace{.1pt}\discretionary{.}{%
}{.}\hspace{.4pt}3\hspace{.1pt}\discretionary{.}{%
}{.}\hspace{.4pt}433}}


\bibitem{reinecke2015labinthewild}
K.~Reinecke and K.~Z. Gajos.
\newblock {LabintheWild: Conducting Large-Scale Online Experiments With
  Uncompensated Samples}.
\newblock In {\em Proceedings of the 18th {ACM} Conference on Computer
  Supported Cooperative Work {\&} Social Computing, {CSCW} 2015, Vancouver, BC,
  Canada, March 14 - 18, 2015}, pp. 1364--1378. {ACM}, 2015. doi: {{%
10\hspace{.1pt}\discretionary{.}{%
}{.}\hspace{.4pt}1145\discretionary{/}{%
}{/}2675133\hspace{.1pt}\discretionary{.}{%
}{.}\hspace{.4pt}2675246}}


\bibitem{renooij1999talking}
S.~Renooij and C.~Witteman.
\newblock {Talking probabilities: Communicating probabilistic information with
  words and numbers}.
\newblock {\em International Journal of Approximate Reasoning}, 22(3):169--194,
  1999. doi: {{%
10\hspace{.1pt}\discretionary{.}{%
}{.}\hspace{.4pt}1016\discretionary{/}{%
}{/}S0888\discretionary{%
}{-}{-}613X\discretionary{%
}{(}{(}99\discretionary{)}{%
}{)}00027\discretionary{%
}{-}{-}4}}


\bibitem{willems2020variability}
C.~A. {Sanne Willems} and I.~S. Verbal.
\newblock {Variability in the Interpretation of Probability Phrases Used in
  Dutch News Articles — a Risk for Miscommunication}.
\newblock {\em Orphanet Journal of Rare Diseases}, 21(1):1--9, 2020.

\bibitem{schapira2001frequency}
M.~M. Schapira, A.~B. Nattinger, and C.~A. McHorney.
\newblock {Frequency or Probability? A Qualitative Study of Risk Communication
  Formats Used in Health Care}.
\newblock {\em Medical Decision Making}, 21(6):459--467, 2001.

\bibitem{spiegelhalter2011visualizing}
D.~Spiegelhalter, M.~Pearson, and I.~Short.
\newblock {Visualizing Uncertainty about the Future}.
\newblock {\em Science}, 333(6048):1393--1400, 2011. doi: {{%
10\hspace{.1pt}\discretionary{.}{%
}{.}\hspace{.4pt}1126\discretionary{/}{%
}{/}science\hspace{.1pt}\discretionary{.}{%
}{.}\hspace{.4pt}1191181}}


\bibitem{sturm2015weird}
C.~Sturm, A.~Oh, S.~Linxen, J.~Abdelnour~Nocera, S.~Dray, and K.~Reinecke.
\newblock {How {WEIRD} is {HCI}? Extending HCI Principles to Other Countries
  and Cultures}.
\newblock In {\em Proceedings of the 33rd Annual ACM Conference Extended
  Abstracts on Human Factors in Computing Systems}, vol.~18, pp. 2425--2428,
  2015. doi: {{%
10\hspace{.1pt}\discretionary{.}{%
}{.}\hspace{.4pt}1145\discretionary{/}{%
}{/}2702613\hspace{.1pt}\discretionary{.}{%
}{.}\hspace{.4pt}2702656}}


\bibitem{talbot2014four}
J.~Talbot, V.~Setlur, and A.~Anand.
\newblock {Four Experiments on the Perception of Bar Charts}.
\newblock {\em IEEE transactions on visualization and computer graphics},
  20(12):2152--2160, 2014.

\bibitem{till2014fostering}
C.~Till.
\newblock {Fostering Risk Literacy in Elementary School}.
\newblock {\em International Electronic Journal of Mathematics Education},
  9(1-2):85--98, 2014.

\bibitem{Visschers2009}
V.~H. Visschers, R.~M. Meertens, W.~W. Passchier, and N.~N. {De Vries}.
\newblock {Probability Information in Risk Communication: A Review of the
  Research Literature}.
\newblock {\em Risk Analysis}, 29(2):267--287, 2009. doi: {{%
10\hspace{.1pt}\discretionary{.}{%
}{.}\hspace{.4pt}1111\discretionary{/}{%
}{/}j\hspace{.1pt}\discretionary{.}{%
}{.}\hspace{.4pt}1539\discretionary{%
}{-}{-}6924\hspace{.1pt}\discretionary{.}{%
}{.}\hspace{.4pt}2008\hspace{.1pt}\discretionary{.}{%
}{.}\hspace{.4pt}01137\hspace{.1pt}\discretionary{.}{%
}{.}\hspace{.4pt}x}}


\bibitem{wallsten1986measuring}
T.~S. Wallsten, D.~V. Budescu, A.~Rapoport, R.~Zwick, and B.~Forsyth.
\newblock {Measuring the Vague Meanings of Probability Terms}.
\newblock {\em Journal of Experimental Psychology: General}, 115(4):348--365,
  1986. doi: {{%
10\hspace{.1pt}\discretionary{.}{%
}{.}\hspace{.4pt}1037\discretionary{/}{%
}{/}0096\discretionary{%
}{-}{-}3445\hspace{.1pt}\discretionary{.}{%
}{.}\hspace{.4pt}115\hspace{.1pt}\discretionary{.}{%
}{.}\hspace{.4pt}4\hspace{.1pt}\discretionary{.}{%
}{.}\hspace{.4pt}348}}


\bibitem{winschiers2013toward}
H.~Winschiers-Theophilus and N.~J. Bidwell.
\newblock {Toward an Afro-Centric Indigenous HCI Paradigm}.
\newblock {\em International Journal of Human-Computer Interaction},
  29(4):243--255, 2013.

\bibitem{wintle2019verbal}
B.~C. Wintle, H.~Fraser, B.~C. Wills, A.~E. Nicholson, and F.~Fidler.
\newblock {Verbal Probabilities: Very Likely to be Somewhat More Confusing than
  Numbers}.
\newblock {\em PLoS ONE}, 14(4):1--18, 2019. doi: {{%
10\hspace{.1pt}\discretionary{.}{%
}{.}\hspace{.4pt}1371\discretionary{/}{%
}{/}journal\hspace{.1pt}\discretionary{.}{%
}{.}\hspace{.4pt}pone\hspace{.1pt}\discretionary{.}{%
}{.}\hspace{.4pt}0213522}}


\bibitem{xiong2022investigating}
C.~Xiong, A.~Sarvghad, D.~G. Goldstein, J.~M. Hofman, and {\c{C}}.~Demiralp.
\newblock Investigating perceptual biases in icon arrays.
\newblock In {\em CHI Conference on Human Factors in Computing Systems}, pp.
  1--12, 2022.

\bibitem{zikmundfisher2014blocks}
B.~J. Zikmund-Fisher, H.~O. Witteman, M.~Dickson, A.~Fuhrel-Forbis, V.~C. Kahn,
  N.~L. Exe, M.~Valerio, L.~G. Holtzman, L.~D. Scherer, and A.~Fagerlin.
\newblock {Blocks, Ovals, or People? Icon Type affects Risk perceptions and
  Recall of Pictographs}.
\newblock {\em Medical Decision Making}, 34(4):443--453, 2014. doi: {{%
10\hspace{.1pt}\discretionary{.}{%
}{.}\hspace{.4pt}1177\discretionary{/}{%
}{/}0272989X13511706}}


\end{thebibliography}
\clearpage\end{CJK}
\end{document}